\documentclass[fleqn,usenatbib]{mnras}

\usepackage[T1]{fontenc}

\DeclareRobustCommand{\VAN}[3]{#2}
\let\VANthebibliography\thebibliography
\def\thebibliography{\DeclareRobustCommand{\VAN}[3]{##3}\VANthebibliography}

\usepackage{graphicx}	
\usepackage{amsmath}	
\usepackage{amssymb}	
\usepackage[referable]{threeparttablex}
\usepackage{multirow}
\usepackage{lipsum}
\usepackage{xcolor}
\usepackage{xspace}

\usepackage{newtxtext,newtxmath}

\newcommand{\msol}{M$_\odot$\xspace}

\newcommand{\VSGR}{{V4641~Sgr}\xspace}
\newcommand{\VCYG}{{V404~Cyg}\xspace}
\newcommand{\SwiftJ}{{Swift~J1357.2$-$0933}\xspace}
\newcommand{\XTEone}{{XTE~J1118$+$480}\xspace}
\newcommand{\XTEtwo}{{XTE~J2012$+$381}\xspace}
\newcommand{\fourU}{{4U~1957$+$115}\xspace}
\newcommand{\maxi}{{MAXI~J1820+070}\xspace}
\newcommand{\maxitwo}{{MAXI~J0637$-$430}\xspace}
\newcommand{\oneA}{{1A~0620$-$00}\xspace}

\newcommand{\mrm}[1]{\mathrm{#1}}
\newcommand{\DUF}{DIPol-UF}
\newcommand{\DP}{DIPol-2}

\newcommand{\dec}[4]{$#1\degr\,#2\arcmin\,#3\farcs#4$}
\newcommand{\ra}[4]{$#1^\mrm{h}\,#2^\mrm{m}\,#3\fs#4$}

\title[Optical polarization of BHXRBs]{Optical polarization signatures of black hole X-ray binaries}

\author[V. Kravtsov et al.]{Vadim Kravtsov$^{1}$\thanks{E-mail: vakrau@utu.fi},
Andrei V. Berdyugin$^{1}$,
Ilia A. Kosenkov$^{1}$,
Alexandra Veledina$^{1,2,3}$,
Vilppu Piirola$^{1}$,
\newauthor
Yasir Abdul Qadir$^{1}$,
Svetlana V. Berdyugina$^{4}$,
Takeshi Sakanoi$^{5}$,
Masato Kagitani$^{5}$ and
Juri Poutanen$^{1,3}$
\\
$^{1}$Department of Physics and Astronomy,  FI-20014 University of Turku, Finland\\
$^{2}$Nordita, KTH Royal Institute of Technology and Stockholm University, Roslagstullsbacken 23, SE-10691 Stockholm, Sweden\\
$^{3}$Space Research Institute of the Russian Academy of Sciences, Profsoyuznaya Str. 84/32, Moscow 117997, Russia\\
$^{4}$Leibniz-Institut f\"{u}r Sonnenphysik, Sch\"{o}neckstr. 6, 79104 Freiburg, Germany\\
$^{5}$Graduate School of Sciences, Tohoku University, Aoba-ku,  980-8578 Sendai, Japan\\
}

\date{Accepted 2022 May 23. Received 2021 April 22; in original form 2021 April 22}

\pubyear{2022}

\begin{document}
\label{firstpage}
\pagerange{\pageref{firstpage}--\pageref{lastpage}}
\maketitle

\begin{abstract}
Polarimetry provides an avenue for probing the geometry and physical mechanisms producing optical radiation in many astrophysical objects, including stellar binary  systems. 
We present the results of multiwavelength (\textit{BVR}) polarimetric studies of a sample of historical black hole X-ray binaries, observed during their outbursts or in the quiescent (or near-quiescent) state. 
We surveyed both long- and short-period systems, located at different Galactic latitudes.  
We performed careful analysis of the interstellar polarization in the direction on the sources to reliably estimate the intrinsic source polarization. 
Intrinsic polarization was found to be small ($<0.2$~per~cent) in sources observed in bright soft states (\maxitwo and \fourU). 
It was found to be significant in the rising hard state of \maxi\ at the level of $\sim 0.5$~per~cent and negligible in the decaying hard state and during its failed outbursts, while \SwiftJ showed its absence in the rising hard state.    
Three (\XTEone, \VSGR, \VCYG) sources observed during quiescence show no evidence of significant intrinsic polarization, while \maxi is the only black hole X-ray binary which showed substantial ($>5$~per~cent) intrinsic quiescent-state polarization with a blue spectrum. 
The absence of intrinsic polarization at the optical wavelengths puts constraints on the potential contribution of non-stellar (jet, hot flow, accretion disc) components to the total spectra of quiescent black hole X-ray binaries. 
\end{abstract}



\section{Introduction}

Accreting stellar-mass black holes in X-ray binaries (BHXRBs) are natural laboratories for studying the interaction between matter and radiation under extreme physical conditions.
During periods of violent activity -- the outbursts -- such systems efficiently convert the gravitational energy into radiation that is observed over a broad range of electromagnetic wavelengths, from radio to X/$\gamma$-rays.
The outburst radio emission is coming from the jet, while the X-rays are produced by the hot accretion flow or corona.
Optical and infrared emission, as evidenced by spectral and timing properties, is a product of a complex interplay between the jet, wind, irradiated disc, and hot accretion flow components \citep{Uttley2014, Poutanen2014a}. 

An outburst typically continues for several weeks to months, eventually decaying into a quiescent state, a long period of inactivity. 
The main components of the system -- the companion star, the accretion disc, the inner accretion flow -- all may contribute to the optical and infrared emission in the quiescence.  
The hot spot/line (the point of intersection of the stream of matter from the companion star and the outer parts of the disc, see e.g., \citealt{McClintock1995, Froning2011}) and jet \citep{Shahbaz2013} were also proposed as potential sources of quiescent emission.

Identifying different spectral components and studying their radiative properties are essential for understanding the mechanisms that trigger the outbursts. 
The contribution of different components to the total spectrum has been studied utilising a variety of methods, with polarimetry often overlooked and undervalued.
Polarization carries information about the geometrical properties of the emitting/scattering media, which may otherwise be inaccessible to an observer.

\begin{table*}
    \centering
    \caption{List of observed BHXRBs.}
    \label{tbl:objects}
    \begin{threeparttable}
        \begin{tabular}{lcccccccc}
            \hline
            \hline
            Object &  Companion & $m_V$ & $\alpha$ & $\delta$ & $\pi$ & $i$ & $P_\text{orb}$ & References\\
                   &             & mag &          &          &  mas  & deg & h & \\
            \hline
            \XTEone  & K7 V -- M1 V & $19.6 \pm 0.2^a$   & \ra{11}{18}{10}{79} & \dec{+48}{02}{12}{32} & $0.30 \pm 0.40$ & $68 \pm 2$ & $4.07841(5)$   &[1, 2, 9]\\
            \SwiftJ  & M5 V         & $17.27 \pm 0.02^a$   & \ra{13}{57}{16}{84} &  \dec{-09}{32}{38}{79} & -- & $>70$ & $2.8 \pm 0.3$  & [1, 3]\\
            \fourU   & --          & $\approx 19.0^b$   & \ra{19}{59}{24}{01} & \dec{+11}{42}{29}{86} & $0.07 \pm 0.15$ & $20 - 70$ & $9.33(1)$  & [1, 4, 5, 10]\\
            \VCYG    & K3 III       & $\approx18.7^c$   & \ra{20}{24}{03}{82} & \dec{+33}{52}{01}{90} & $ 0.42 \pm 0.02 $ & $67 \pm 3$ & $155.35(2)$  & [1, 6, 7, 11]\\ 
            \VSGR    & B9 III       & $\approx13.5^c$ & \ra{18}{19}{21}{63} & \dec{-25}{24}{25}{85} & $ 0.17 \pm 0.03 $ & $72 \pm 4$ & $67.61(2)$  & [1, 8, 12]\\ 
            \XTEtwo    & --       & $21.3\pm0.1^d$ & \ra{20}{12}{37}{76} & \dec{+38}{11}{00}{77} &  -- & -- & -- & [1]\\
            \maxi      &  K6 IV   & -- & \ra{18}{20}{21}{94} & \dec{+07}{11}{07}{29} & $ 0.37 \pm 0.08 $ & $73 \pm 6$ & $16.4518(2)$ & [1, 13, 14, 15]\\ 
            \maxitwo   & --     & $\approx16.5^c$ & \ra{06}{36}{23}{59} & \dec{-42}{52}{04}{10} & -- & -- & -- & [1]\\
            \hline
        \end{tabular}
        \begin{tablenotes}
        \item
        References: (1) \citet{GaiaDR3}, (2) \citet{Gelino2006}, (3) \citet{CorralSantana2013}, (4) \citet{Hakala1999}, (5) \citet{Bayless2011}, (6) \citet{Miller-Jones2009}, (7) \citet{Khargharia2010}, (8) \citet{MacDonald2014}, (9) \citet{Torres2004}, (10) \citet{Thorstensen1987}, (11) \citet{Casares1992}, (12) \citet{Orocz2011}, (13)  \citet{Torres2020}, (14) \citet{Poutanen2022}, (15) \citet{Mikolajewska22}.\\
        $^a$StanCam photometry, $^b$\citet{Hakala2014}, $^c$AAVSO magnitudes, $^d$\citet{Hynes1999}.
        \end{tablenotes}
    \end{threeparttable}
\end{table*}

\begin{figure*}
    \centering
    \includegraphics[trim={2cm 0 2cm 0}, clip, width=0.43\linewidth]{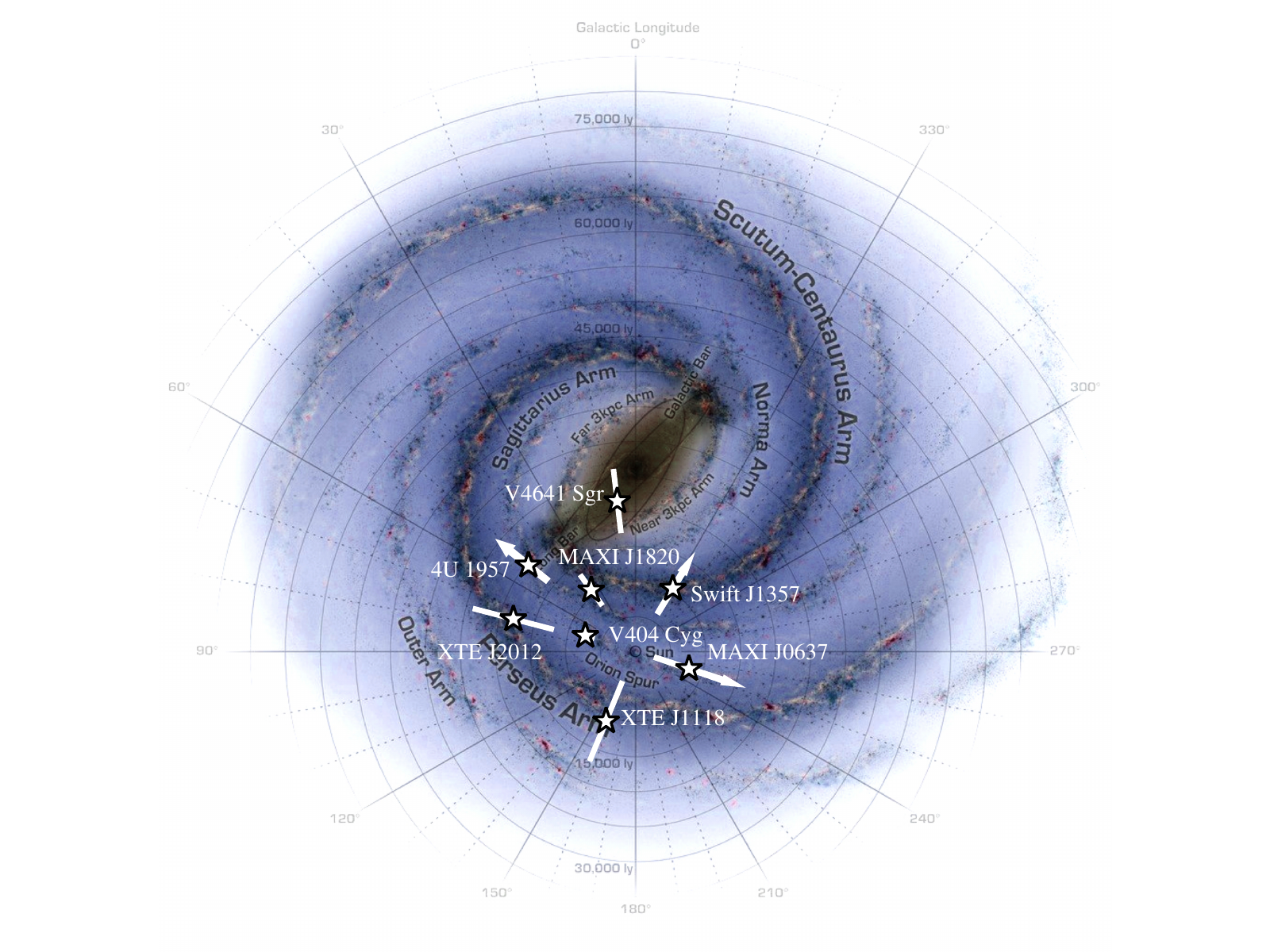}
    \hspace{0.5cm}
    \includegraphics[width=0.53\linewidth]{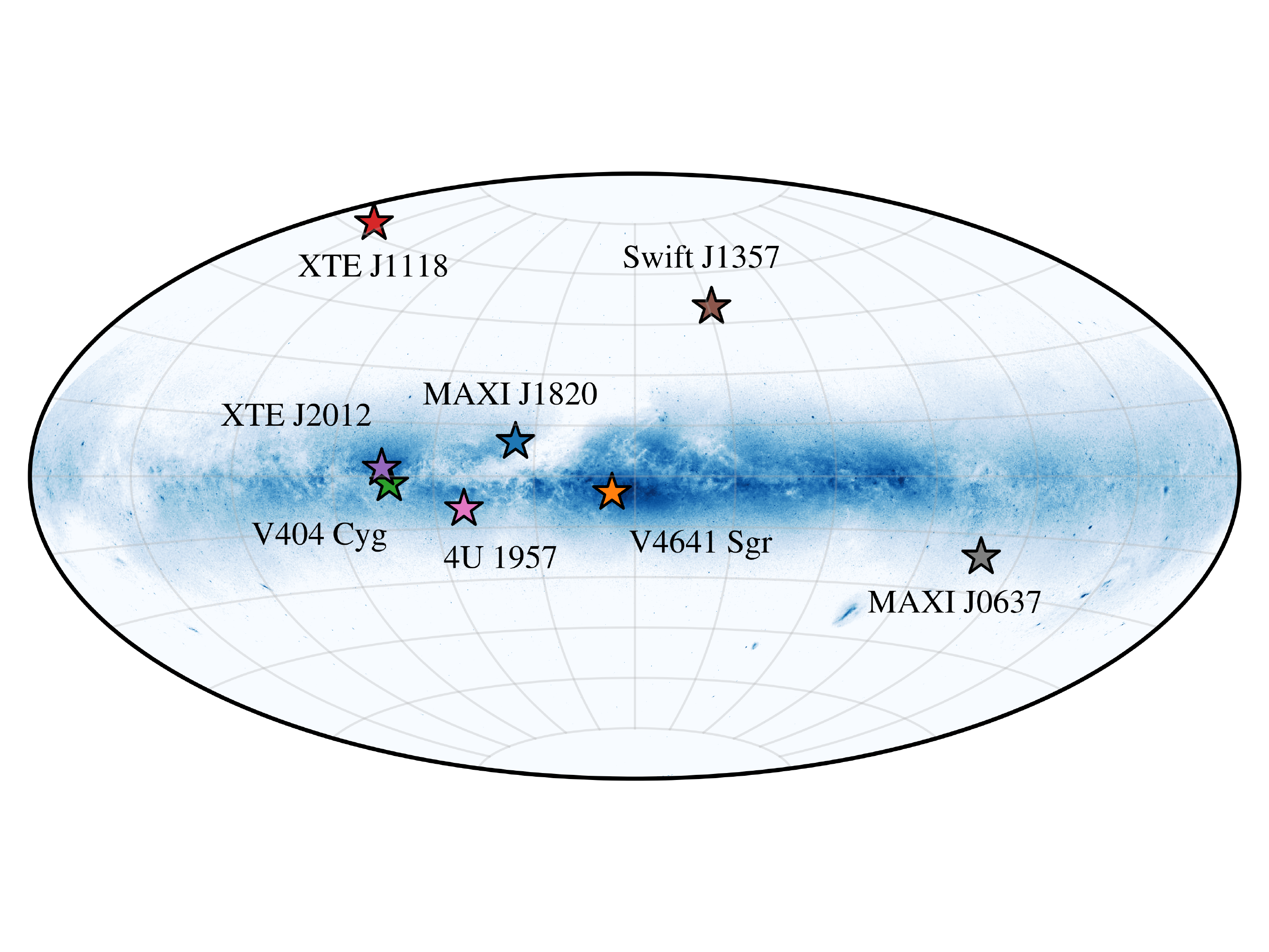}
    \caption{Galactic distribution of the observed X-ray binaries: pole-on view (left) and edge-on view (right). Background image credit: NASA/JPL-Caltech/R.Hurt (SSC/Caltech), inverted.}
    \label{fig:galmap}
\end{figure*}

Polarized radiation can be produced by several physical processes, including synchrotron radiation in the presence of an ordered magnetic field of the jet or hot accretion flow, electron scattering in the accretion disc atmosphere or scattering of the accretion disc radiation in the outflow (jet/wind) by electrons or, in quiescence, by dust.
Each component has different polarization characteristics (or no polarization at all), which makes polarimetry an excellent tool for probing the geometry of the source and physical mechanisms responsible for optical radiation in black hole X-ray binaries. 

A number of recent studies of polarized optical emission of black hole X-ray binaries were focused on BHXRBs in the outbursts \citep[e.g.,][]{Tanaka2016, Shahbaz2016, Itoh2017, Kosenkov2017, Veledina2019, Kosenkov2020b}. 
However, only a few attempts to study quiescent polarization have been made to date.
\citet{Dolan1989} found variable optical polarization of BHXRB \oneA and constrained its inclination by modelling the dependence of Stokes parameters on orbital phase. 
Significant quiescent optical polarization of 1A~0620$-$00 at the level of $P \approx 3$~per~cent was later confirmed \citep{Dubus2008}.
\citet{Russell2016} claimed detection of near-infrared quiescent intrinsic polarization of \oneA and \SwiftJ.
\maxi, observed in near-quiescence after its 2018 outburst, demonstrated a large intrinsic polarization degree exceeding 5 per cent in \textit{B}-band and blue polarization spectra, likely caused by scattering either in the hot accretion flow or by the dusty equatorial wedge \citep{Poutanen2022}. 
A difference by 45\degr\ between polarization angle and the position angle of the jet was interpreted as a signature of a high, more than 40\degr, misalignment between the orbital angular momentum and the black-hole spin \citep{Poutanen2022}. 
Similar polarization signatures, if found in other sources, could be used to study the statistical distribution of orbital-BH spin misalignment angle, constraining binary evolution and black hole formation scenarios.

In this paper, we present a study of optical ($BVR$) polarization of a sample of BHXRBs during the outbursts and in the quiescent or near-quiescent states.
We surveyed both long- and short-period systems with different spectral classes of the companion star, located at different Galactic latitudes. 
We put tight constraints on the magnitude of intrinsic polarization for most of the sources with the help of polarimetric observations of the field stars.
The properties of intrinsic polarization allowed us to estimate the potential contribution of non-stellar components (jet/hot flow) to the total spectra.

\begin{table*}
    \centering
    \caption{Observed polarization of BHXRBs. Errors are $1\sigma$.}
    \label{tbl:pol}
    \begin{tabular}{lcc*6{r@{\,$\pm$\,}l}} 
        \hline
        \hline
        & & & \multicolumn{4}{c}{$B$} & \multicolumn{4}{c}{$V$} & \multicolumn{4}{c}{$R$} \\
        Instrument & State & Date & \multicolumn{2}{c}{$P$} & \multicolumn{2}{c}{$\theta$} & \multicolumn{2}{c}{$P$} & \multicolumn{2}{c}{$\theta$} & \multicolumn{2}{c}{$P$} & \multicolumn{2}{c}{$\theta$} \\
        & & MJD &  \multicolumn{2}{c}{\%} & \multicolumn{2}{c}{deg} &  \multicolumn{2}{c}{\%} & \multicolumn{2}{c}{deg} &  \multicolumn{2}{c}{\%} & \multicolumn{2}{c}{deg} \\
        \hline
        & & &\multicolumn{7}{c}{\XTEone} \\
        \DUF  & Quiescence & 59326.03 & 1.65 & 0.80 & 22.7 & 13.0 & 1.76 & 0.77 & 79.4 & 11.8 & 1.49 & 0.53 & 51.2 & 9.9 \\
        \hline
        & & & \multicolumn{7}{c}{\SwiftJ} \\
        \DUF  & Failed outburst & 59326.10 & 0.34 & 0.07 & 34.5 & 6.1  & 0.18 & 0.08 & 4.2  & 12.4 & 0.25 & 0.06 & 69.8 & 8.5 \\
        \hline
        & & & \multicolumn{7}{c}{\fourU} \\
        \DUF  & Soft & 59401.14 & 0.65 & 0.07 & 59.4 & 3.1  & 0.61 & 0.09 & 54.8 & 4.0  & 0.62 & 0.08 & 60.2 & 3.9 \\
        \hline
        & & & \multicolumn{7}{c}{\VSGR} \\
        \multirow{3}{*}{\DP} &  & 58347.40 & 0.46 & 0.05 & 34.6 & 2.9 & 0.52 & 0.07 & 34.8 & 4.0 & 0.54 & 0.07 & 40.6 & 3.5 \\
                             & Quiescence & 58348.40 & 0.41 & 0.04 & 44.7 & 2.8 & 0.38 & 0.05 & 38.7 & 3.9 & 0.39 & 0.04 & 44.3 & 2.7 \\
                             &  & 58351.39 & 0.42 & 0.08 & 33.0 & 5.0 & 0.47 & 0.08 & 29.8 & 5.0 & 0.50 & 0.09 & 28.4 & 5.0 \\
        \noalign{\vskip 2mm}
        \multirow{5}{*}{\DUF} &  & 58686.98 & 0.53 & 0.03 & 43.6 & 1.6 & 0.40 & 0.04 & 46.0 & 2.7 & 0.45 & 0.02 & 50.7 & 0.9 \\
                              &  & 58961.19 & 0.46 & 0.02 & 36.8 & 1.4 & 0.49 & 0.06 & 45.7 & 3.6 & 0.42 & 0.06 & 48.7 & 4.3 \\ 
                              & Quiescence & 58964.21 & 0.49 & 0.01 & 40.2 & 0.8 & 0.43 & 0.03 & 44.7 & 1.8 & 0.44 & 0.02 & 50.7 & 1.0 \\
                              &  & 58966.22 & 0.50 & 0.03 & 40.1 & 1.4 & 0.44 & 0.04 & 46.0 & 2.3 & 0.48 & 0.04 & 50.0 & 2.4 \\
                              &  & 58967.20 & 0.51 & 0.05 & 44.4 & 2.9 & 0.50 & 0.10 & 43.2 & 5.7 & 0.45 & 0.10 & 51.2 & 6.1 \\
        \noalign{\vskip 2mm}
        \multirow{3}{*}{\DP} &  & 59519.71 & 0.56 & 0.08 & 38.5 & 4.0 & 0.36 & 0.15 & 31.9 & 10.9 & 0.29 & 0.12 & 46.9 & 11.2 \\
                             & Failed outburst & 59521.71 & 0.54 & 0.08 & 41.8 & 4.2 & 0.37 & 0.20 & 45.7 & 13.9 & 0.58 & 0.09 & 49.5 & 4.6 \\
                             &  & 59522.71 & 0.47 & 0.08 & 42.1 & 4.8 & 0.48 & 0.48 & 38.6 & 4.4  & 0.50 & 0.10 & 44.4 & 5.6 \\
        
        \hline
        & & & \multicolumn{7}{c}{\VCYG} \\
        \DP & Rising hard$^a$  &  57195--57200 & 8.55 & 0.20 & 6.7 & 0.7 & 7.47  & 0.06 & 8.6 & 0.2  & 7.51  & 0.03 & 6.8 & 0.1 \\
        \noalign{\vskip 2mm}
        \DP & Quiescence$^a$ &  57206--57210 & 7.84 & 0.16 & 7.9 & 0.6 & 6.58  & 0.05 & 11.1 & 0.2  & 7.13  & 0.03 & 7.7 & 0.1 \\
        \noalign{\vskip 2mm}
        \DP & Quiescence$^a$ &  57651--57652 & \multicolumn{2}{c}{--} & \multicolumn{2}{c}{--} & 7.32  & 0.38 & 9.7 & 1.5  & 7.37  & 0.21 & 7.2 & 0.8 \\
        \noalign{\vskip 2mm}
        \multirow{2}{*}{\DUF}  & \multirow{2}{*}{Quiescence} & 58688.02 & \multicolumn{2}{c}{--}   & \multicolumn{2}{c}{--} & 8.07 & 0.41 & 3.1 & 1.5 & 7.30 & 0.14 & 5.3 & 0.5 \\
        &  & 59402.16 & 7.15 & 0.47 & 3.6  & 1.9  & 7.78 & 0.21 & 5.5  & 0.8  & 7.85 & 0.07 &  6.9 & 0.3 \\
        
        \hline
        & & & \multicolumn{7}{c}{\maxi} \\
        \multirow{2}{*}{\DP} & \multirow{2}{*}{Rising hard$^b$}      & 58195--58222 & 0.76 & 0.01 & 53.9  & 0.3  & 0.79 & 0.01 & 54.7  & 0.4  & 0.76 & 0.01 &  53.3 & 0.3 \\
                             &                                   & 58222--58234 & 0.76 & 0.02 & 51.4  & 0.6  & 0.87 & 0.02 & 50.5  & 0.8  & 0.86 & 0.02 &  45.8 & 0.6 \\
        \noalign{\vskip 2mm}
        \DP                    &  Soft$^c$          & 58312--58344 & 0.66 & 0.01 & 61.5  & 0.4  & 0.67 & 0.01 & 62.2  & 0.5  & 0.62 & 0.01 &  63.5 & 0.4 \\
        \noalign{\vskip 2mm}
        \DP                  &  Decaying hard$^c$   & 58406--58428 & 0.76 & 0.04 & 62.2  & 1.4  & 0.63 & 0.06 & 64.1  & 2.6  & 0.67 & 0.04 &  62.0 & 1.7 \\
        \noalign{\vskip 2mm}
        \multirow{2}{*}{\DP} & \multirow{2}{*}{Failed outburst} & 58721--58726 & 0.67 & 0.13 & 65.9 & 5.6 & 0.78 & 0.15 & 65.0 & 5.5 & 0.62 & 0.10 & 64.3 & 4.7 \\
                             & & 58911--58932 & 0.60 & 0.11 & 64.5 & 5.3 & 0.79 & 0.20 & 65.4 & 7.5 & 0.71 & 0.12 & 68.4 & 5.0 \\
        
        \noalign{\vskip 2mm}
         \DUF                & Quiescence$^d$ & 58961--59401 & 2.88 & 0.26 & $-$16.8 & 3.1 & 1.67 & 0.25 & $-$13.6 & 5.5 & 0.63 & 0.17 & 1.5 & 7.0\\
        
        \hline
        & & & \multicolumn{7}{c}{\maxitwo} \\
        \DP & Soft & 58792--58796  &  0.28 & 0.26 & 38.8  & 26.5  & \multicolumn{2}{c}{--} & \multicolumn{2}{c}{--} & \multicolumn{2}{c}{--} & \multicolumn{2}{c}{--} \\
        
        \hline
    \end{tabular}
    \begin{tablenotes}
            \item Source of the data:  $^a$\citet{Kosenkov2017}, $^b$\citet{Veledina2019}, $^c$\citet{Kosenkov2020b}, $^d$\citet{Poutanen2022}.
    \end{tablenotes}
\end{table*}

\begin{table*}
    \centering
    \caption{Polarization of field stars.}
    \label{tbl:ref_pol}
    \begin{tabular}{lccc*6{r@{\,$\pm$\,}l}} 
        \hline
        \hline
        & & & & \multicolumn{4}{c}{$B$} & \multicolumn{4}{c}{$V$} & \multicolumn{4}{c}{$R$} \\
        Field star & Identifier & Parallax & Angular & \multicolumn{2}{c}{$P$} & \multicolumn{2}{c}{$\theta$} & \multicolumn{2}{c}{$P$} & \multicolumn{2}{c}{$\theta$} & \multicolumn{2}{c}{$P$} & \multicolumn{2}{c}{$\theta$} \\
        & HD/BD/Gaia DR3 & mas & separation &  \multicolumn{2}{c}{\%} & \multicolumn{2}{c}{deg} &  \multicolumn{2}{c}{\%} & \multicolumn{2}{c}{deg} &  \multicolumn{2}{c}{\%} & \multicolumn{2}{c}{deg} \\
        \hline
        & &  \multicolumn{9}{c}{\XTEone} \\
        Ref A$^a$  & BD+48~1955 & $4.18\pm0.02$  & $\sim100^\prime$ & \multicolumn{2}{c}{--} & \multicolumn{2}{c}{--} & 0.06 & 0.03 & 89 & 12 & \multicolumn{2}{c}{--} & \multicolumn{2}{c}{--} \\
        \hline
        & & \multicolumn{9}{c}{\SwiftJ} \\
        Ref A$^a$  & HD~122835 & $4.15\pm0.30$ & $\sim100^\prime$ & \multicolumn{2}{c}{--} & \multicolumn{2}{c}{--}  & 0.12 & 0.05 & 101  & 13 & \multicolumn{2}{c}{--} & \multicolumn{2}{c}{--} \\
        \hline
        & &\multicolumn{9}{c}{\fourU} \\
        Ref A  & 4303869526257087360 & $0.36\pm0.13$ & $<1^\prime$ & 0.59 & 0.15 & 53 & 7 & 0.72 & 0.12 & 53 & 5  & 0.48 & 0.08 & 62 & 5 \\
        Ref B  & 4303869599285320832 & $0.17\pm0.08$ & $<1^\prime$ & 0.65 & 0.07 & 55 & 3 & \multicolumn{2}{c}{--} & \multicolumn{2}{c}{--} & 0.52 & 0.05 & 54 & 3 \\
        \hline
        & &\multicolumn{9}{c}{\VSGR} \\
        Ref A  & 4053096384526868736 & $0.24\pm0.03$ & $<10^\prime$ & 0.56 & 0.08 & 55 & 4 & 0.40 & 0.03 & 50 & 2 & 0.39 & 0.02 & 56 & 1 \\
        Ref B  & 4053096491998429952 & $0.40\pm0.03$ & $<10^\prime$ & 0.54 & 0.05 & 52 & 3 & 0.62 & 0.08 & 51 & 4 & 0.59 & 0.06 & 62 & 3 \\
        Ref C  & 4053096315807371008 & $0.51\pm0.03$ & $<10^\prime$ & 0.26 & 0.05 & 59 & 5 & 0.29 & 0.04 & 62 & 4 & 0.33 & 0.02 & 63 & 2 \\
        Ref D  & 4053096320199414528 & $0.65\pm0.03$ & $<10^\prime$ & 0.57 & 0.08 & 67 & 4 & 0.45 & 0.09 & 65 & 6 & 0.45 & 0.06 & 74 & 4 \\
        Ref E  & 4053096595077613568 & $0.42\pm0.03$ & $<10^\prime$ & 0.70 & 0.11 & 56 & 4 & 0.47 & 0.07 & 56 & 4 & 0.40 & 0.02 & 51 & 2 \\
        Ref F  & 4053096487606085632 & $0.53\pm0.03$ & $<10^\prime$ & 0.29 & 0.08 & 67 & 8 & 0.38 & 0.08 & 77 & 6 & 0.35 & 0.06 & 69 & 5 \\
        
        \hline
        \noalign{\vskip 2mm}
        & &\multicolumn{9}{c}{\VCYG} \\
        Ref 4040$^b$  & 2056188620566335360 & $0.14\pm0.11$ & $1.4^{\prime\prime}$ & \multicolumn{2}{c}{--} & \multicolumn{2}{c}{--}  & 6.64 & 0.22 & 12 & 1  & 7.28 & 0.09 & 9 & 1 \\
        Ref 4042$^b$  & 2056188865390747136 & $0.35\pm0.04$ & $<10^\prime$ & \multicolumn{2}{c}{--} & \multicolumn{2}{c}{--}  & 7.09 & 0.42 & 11 & 2  & 8.47 & 0.17 & 9 & 1 \\
        Ref 4043$^b$  & 2056190136700843264 & $0.34\pm0.03$ & $<10^\prime$ & 6.92 & 0.29 & 3 & 1 & 5.20 & 0.15 & 11 & 1  & 6.47 & 0.07 & 11 & 1 \\
        
        \hline
        & &\multicolumn{9}{c}{\XTEtwo} \\
        Ref A  & 2061667766205170048 & $0.23\pm0.05$ & $<1^\prime$ & 3.36 & 0.60 & 88 & 5 & 3.90 & 0.19 & 84 & 1  & 3.69 & 0.07 & 83 & 1 \\
        Ref B  & 2061673435561995008 & $0.19\pm0.05$ & $<1^\prime$ & 2.58 & 0.42 & 83 & 5 & 3.78 & 0.16 & 90 & 1  & 3.71 & 0.07 & 88 & 1\\
        \hline
        
        & &\multicolumn{9}{c}{MAXI~J1820+070} \\
        Ref 1-5  & 2,3,6,7,9$^c$ & $0.15 - 0.44$ & $<10^\prime$ & $0.80$ & $0.03$ & $64$ & $1$ & $0.70$ & $0.03$ & $69$ & $1$ & $0.60$ & $0.02$ & $64$ & $1$ \\

        \hline
        
        & &\multicolumn{9}{c}{\maxitwo} \\
        Ref A  & 5569292377717900288 & $0.67\pm0.01$ & $<10^\prime$ & $0.57$ & $0.08$ & $39$ & $4$ & \multicolumn{2}{c}{--} & \multicolumn{2}{c}{--} & \multicolumn{2}{c}{--} & \multicolumn{2}{c}{--} \\
        Ref B  & 5569291931041304960 & $0.78\pm0.02$ & $<10^\prime$ & $0.42$ & $0.08$ & $38$ & $6$ & \multicolumn{2}{c}{--} & \multicolumn{2}{c}{--} & \multicolumn{2}{c}{--} & \multicolumn{2}{c}{--} \\
        Ref C  & 5569291999760781312 & $1.09\pm0.26$ & $<10^\prime$ & $0.14$ & $0.07$ & $67$ & $14$& \multicolumn{2}{c}{--} & \multicolumn{2}{c}{--} & \multicolumn{2}{c}{--} & \multicolumn{2}{c}{--} \\
        
        \hline
       
    \end{tabular}
    \begin{tablenotes}
            \item $^a$Nearest stars from the catalogue of \citet{Berdyugin2014}.
            \item $^b$Reference stars from Table 3 of \citet{Kosenkov2017}.
            \item $^c$Polarization is given as the weighted average of the polarization of five field stars from Table 2 of \citet{Veledina2019}.
    \end{tablenotes}
\end{table*}

\section{Data Acquisition and Reduction}
\label{seq:data}

The observations of BHXRBs were carried out with two copies of \DP\ polarimeter \citep{Piirola2014} and a single unit of \DUF\ polarimeter \citep{Piirola2020a}. 
One copy of \DP\ is installed at the 60~cm Tohoku T60 telescope, Haleakala Observatory, Hawaii, USA; another was mounted on the 60~cm KVA telescope, Observatorio del Roque de los Muchachos (ORM), La Palma, Spain, and was also used as a visitor instrument at 4.2~m William Herschel Telescope (WHT, ORM) and 2.2~m Univesity of Hawaii telescope (UH88, Mauna Kea, Hawaii, USA).
\DUF\ is a visitor instrument installed at the 2.56~m Nordic Optical Telescope (NOT, ORM).

We have collected the polarimetric data on eight BHXRBs with declination $\delta>-30\degr$ (constrained by the geography of the telescopes used) and visual magnitude $m_V < 21$~mag, limited by the capabilities of the NOT (see Table~\ref{tbl:objects}).
These include both short- and long-period systems ($P_\text{orb}$ from $\sim 2.8$~hours to 6.5~days), systems with different spectral classes of companion stars (from B9~III to M5~V), located at broad range of Galactic latitudes (Fig.~\ref{fig:galmap} and Table~\ref{tbl:objects}).

\XTEone, \SwiftJ, \fourU and \XTEtwo were observed for one night each (\DUF\ at NOT). \maxitwo was observed for three nights during its soft state in November 2019 (\DP\ at T60).
\VSGR was observed for a total of 11 nights: eight nights during its quiescent state (three nights with \DP\ at T60 in 2018 and five nights with \DUF\ at NOT in 2019--2020) and for another three nights during its 2021 failed outburst (\DP\ at T60). 
\VCYG was observed for two nights during its quiescent state in July 2019 and July 2021 (\DUF\ at NOT).
\maxi was observed for a total of 10 nights during its failed outbursts in August 2019 and March 2020 (\DP\ at T60).

We complement the new measurements with the previously published \DP\ and \DUF\ data. These data include the results of polarimetric observations of \VCYG and \maxi during a total of 12 and 65 nights respectively. \VCYG was monitored for five night during its 2015 outburst (\DP\ at T60), for another five nights after the outburst has ended (\DP\ at WHT) and for two nights during the quiescent state with \DP\ mounted at the UH88 \citep{Kosenkov2017}. \maxi was monitored for 12 nights during the rising hard state (\DP\ at T60, \citealt{Veledina2019}), for 26 and 9 nights during the soft and decaying hard states respectively (\DP\ at T60, \citealt{Kosenkov2020b}), and for 18 nights during the quiescent state (\DUF\ at NOT, \citealt{Poutanen2022}).  

The only relatively bright source in the sample, \VSGR, was observed using a conventional amplifier, while for fainter targets we used electron-multiplication regime of \DUF, which provides better signal-to-noise ratio under such conditions \citep[for a detailed description of instrument modes, see ][]{Piirola2020a,KosenkovPHD}. 
In addition to the polarimetry, we were able to perform photometric measurements of \XTEone and \SwiftJ with the StanCam CCD photometer, mounted at the NOT. 
The photometric observations were made within the same night as the polarimetric measurements (MJD 59326).
We also used the public AAVSO light curves to estimate \textit{V}-band magnitudes of \VCYG,  \VSGR, and \maxitwo. Stellar magnitudes for all targets are given in  Table~\ref{tbl:objects}.

Both instruments used for polarimetric observations are remotely-operated \citep{DUF_Software} `double-image' CCD polarimeters capable of obtaining polarization images in three ($BVR$) filters simultaneously.
The optical beam from a star is split into two orthogonally polarized rays (ordinary `o' and extraordinary `e'), resulting in two separate and orthogonally polarized images of a star recorded in different parts of the CCD sensor.
The orthogonally polarized images of the sky overlap on each stellar image, effectively eliminating the sky polarization at the instrument level.
The accuracy of polarization measurements can reach $10^{-5}$, limited in practice by the photon noise \citep{Piirola1973, Berdyugin2018, Piirola2020}.

Each obtained image undergoes standard calibration procedures, including bias and dark subtraction and flat fielding \citep{Berdyugin2019}.
The difference in brightness between `o' and `e' images is measured using differential aperture photometry, and Stokes parameters are computed from their intensity ratios.
The individual Stokes parameters are then averaged using `2$\sigma$' averaging procedure \citep{Piirola_Thesis, KosenkovPHD, TwoSigmaWeighting}, obtaining average Stokes parameters and their statistical errors, which are then used to calculate average polarization degree and polarization angle \citep{Simmons1985}. 

The presence of an interstellar (IS) medium between the observer and the object affects the observed polarization. 
The IS polarization has to be estimated and subtracted from the observed polarization. 
One of the most reliable methods for estimating the IS polarization component is to observe a sample of field stars located at distances similar to that of the source. 
For each source at low galactic latitudes, we observed at least two field stars with close parallaxes, while for the high-latitude objects we used the data from the catalogue of \citet{Berdyugin2014}.

\section{Results}

Results of optical polarization measurements obtained for our sample of BHXRBs are given in Table \ref{tbl:pol}. The results of the determination of their IS polarization are shown in Table \ref{tbl:ref_pol}. The intrinsic polarization estimates are given in Table \ref{tab:pol_intr}.

\begin{figure*}
    \centering
    \includegraphics[width=0.95\linewidth]{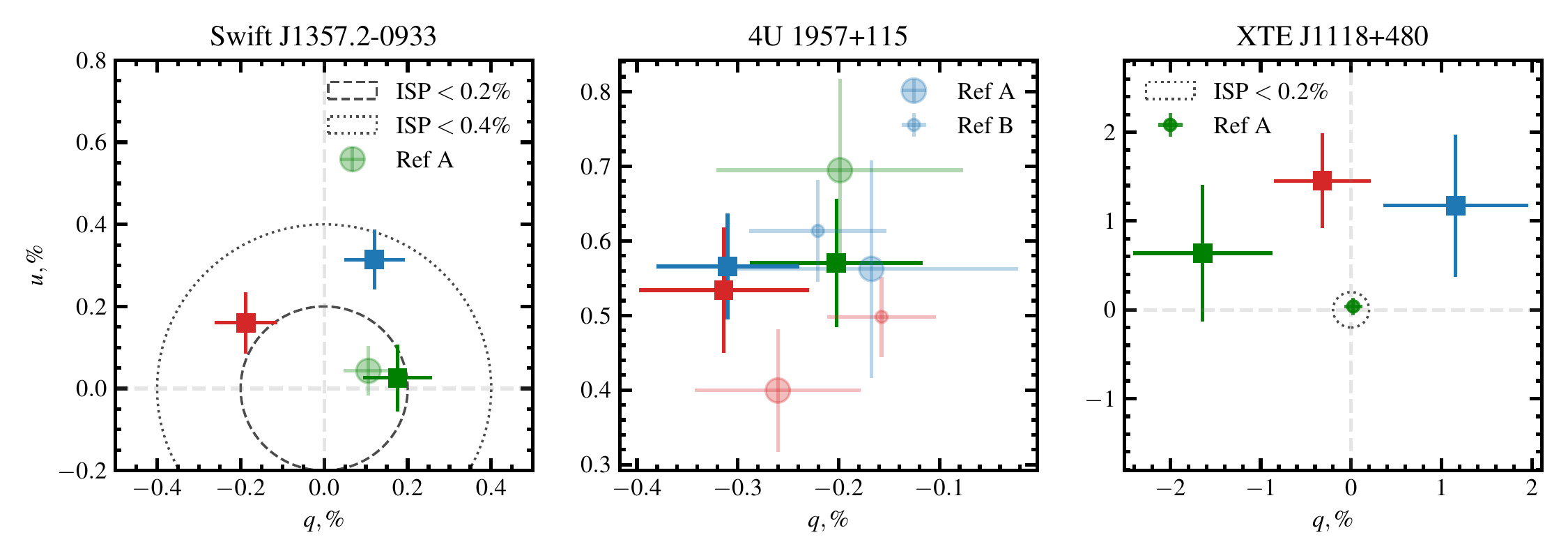}
    \caption{Normalized observed Stokes parameters ($q,u$) for \SwiftJ, \fourU, and \XTEone (from left to the right). The blue, green, and red squares with $1\sigma$ errors correspond to the $B$, $V$, and $R$ optical polarimetric measurements of the targets and the circles correspond to nearby stars.}
    \label{fig:quplane_pannel}
\end{figure*}

\subsection{XTE~J1118+480}

\XTEone\ was discovered during its outburst in 2000 by the {\it Rossi X-ray Timing Explorer} All-Sky Monitor \citep[{\it RXTE}/ASM,][]{Remillard2000}.
The high galactic latitude ($b \approx 62 \degr)$ and large distance from the Galactic plane ($ \approx $ 1.7~kpc) result in a very small absorption in the direction of the source, making it one of the most popular laboratories for studying outbursts and quiescent states in BHXRBs. 
The mass of the black hole is estimated to be 6--9~\msol \citep{Wagner2001, Gelino2006, Chatterjee2019}, while the mass of the companion star is $0.3  \pm 0.2$~\msol \citep{Mirabel2001}.
The orbital period of this system is $P_\text{orb} \approx 4$~h \citep{Torres2004}.
The second outburst in 2005 was extensively monitored in different wavelengths from the radio and optical to the X-rays \citep{Pooley2005, Rupen2005, Remillard2005, Zurita2006}. 

Because of the high galactic latitude,  the stellar number density in the direction of \XTEone is relatively small. As a result, there are no stars located within the instrument field of view ($\sim 1\arcmin$ in the \textit{B}-band and $\sim 45\arcsec$
in the \textit{V} and \textit{R}-bands).
Fortunately, the IS medium density decreases dramatically at high galactic latitudes, making the contribution of the IS polarization component negligible.
IS polarization survey of the high galactic latitudes \citep{Berdyugin2014}, puts an upper limit on the IS polarization in the direction of \XTEone\  of $P_\text{IS} < 0.2$~per~cent (Table~\ref{tbl:ref_pol} and Fig.~\ref{fig:quplane_pannel}).
To check if the polarization of \XTEone differs significantly from this value, we first need to correct its observed polarization degree for the bias, which arises due to the small  signal-to-noise ratio ($P/\sigma < 5$), shifting the polarization degree towards higher values. 
The unbiased maximum likelihood estimator $P_0 = (P^2 - 2 \sigma^2)^{1/2}$ from \citet{Simmons1985} gives us the following estimations of the true values of the polarization degree: $P_{B,0} = 1.2 \pm 0.8$, $P_{V,0} = 1.4 \pm 0.8$ and $P_{R,0} = 1.3 \pm 0.5$~per~cent.
Based of these data we can only confirm the absence of substantial (e.g. $\ge 4$ per cent) optical intrinsic polarization in the quiescence for this transient.
A previous polarimetric measurement of $P_\text{obs} = 0.21 \pm 0.16$~per~cent in the \textit{V}-band during the outburst in 2000 is consistent with the IS polarization level \citep{Schultz2004}.

\subsection{Swift~J1357.2--0933}

The black hole transient \SwiftJ was discovered in 2011 using {\it Neil Gehrels Swift Observatory} Burst Alert Telescope \citep[{\it Swift}/BAT,][]{Krimm2011}. 
Similar to \XTEone, the binary separation of \SwiftJ is very small (the orbital period $P_\text{orb} \approx 2.8~\text{h}$) and the black hole in the system has a mass $M_\text{BH} > 9$~\msol \citep{MataSanchez2015, Casares2016}. 
The analysis of optical spectra revealed remarkable broad double-peaked $\text{H}\alpha$ emission line, which is a strong indication of a high ($i > 70\degr$) binary inclination \citep{CorralSantana2013}.   

During our observations of the source, the beginning of its optical and X-ray re-brightening event was reported \citep{ATel14539, ATel14573, ATel14623, ATel14729}.  
There are no nearby stars in the field of view of \SwiftJ, but its location at the high galactic latitude ($\approx 50\degr$) allows us to put the upper limit of 0.2 per cent on the expected level of IS polarization from the survey of \citet{Berdyugin2014},  see Table~\ref{tbl:ref_pol} and Fig.~\ref{fig:quplane_pannel}. 
The small value of the observed polarization of the source is consistent with the IS polarization level and hence the optical polarization of \SwiftJ\ observed during its transition from the quiescence to the faint outburst (with X-ray luminosity of about $L_{\rm X} \sim 10^{34}$~erg\,s$^{-1}$; \citealt{ATel14573}), most likely has an IS origin.

\citet{Shahbaz2003} argued that the quiescent optical to mid-infrared emission is dominated by the synchrotron jet emission.
This emission is expected to be strongly (up to 70 per cent) polarized, which allows us to estimate its contribution to the total optical spectrum. 
Our non-detection of intrinsic polarization at the level of $P_\text{int} < 0.2$~per~cent suggests this contribution to be less than a few percent of the total optical emission during the initial rise to the outburst. 
Additional polarization measurements during the true quiescent state are needed to estimate the role of the synchrotron emission to the quiescent optical spectrum of \SwiftJ.

\subsection{4U~1957+115}

\fourU was detected by \textit{Uhuru} satellite in 1973 \citep{Giacconi1974} and since then remains in the soft state. 
Its emission is dominated by the accretion disc \citep{Wijnands2002} and modulated with the orbital period $P_\text{orb} \approx 9.3~\text{h}$ in optical light \citep{Thorstensen1987, Hakala1999, Bayless2011, Hakala2014}, while the X-rays show no orbital modulation \citep{Nowak1999}.
Optical light curve modelling \citep{Bayless2011} constrained inclination to be in range of $20\degr < i < 70\degr$. 

To estimate the IS polarization in the direction of \fourU, we measured the polarization of two nearby field stars (Table \ref{tbl:ref_pol} and Fig.~\ref{fig:quplane_pannel}). The observed degree $P_\text{obs} = 0.60\pm 0.08$ per cent and the position angle $\theta_\text{obs} = 58\degr \pm 5\degr$  of linear polarization of \fourU coincides with the IS values within the measurement errors.

\begin{figure}
    \centering
    \includegraphics[width=0.95\linewidth]{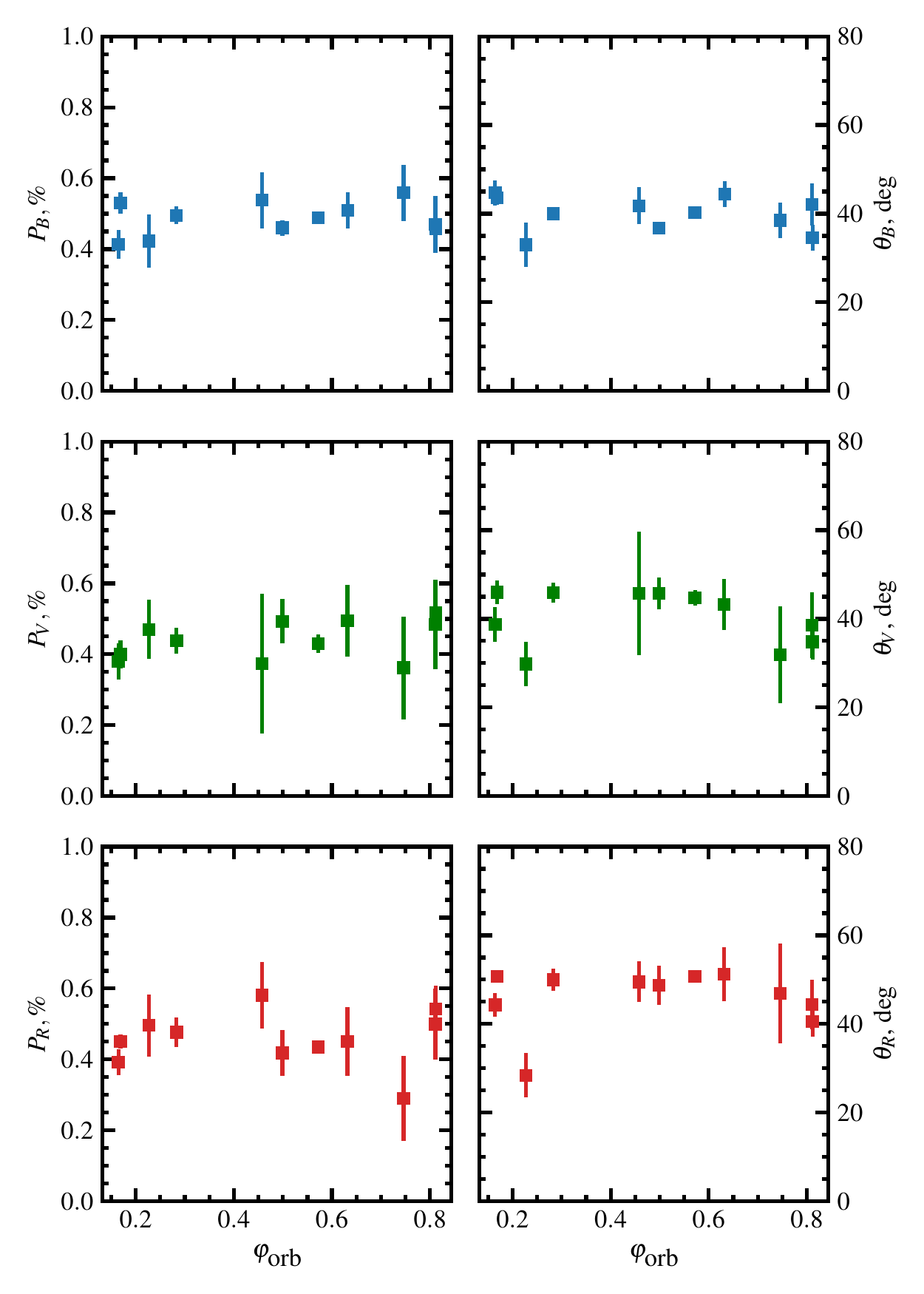}
    \caption{Dependence of the observed polarization degree (\textit{left column}) and polarization angle (\textit{right column}) of \VSGR on the orbital phase in the $BVR$ bands (\textit{from top to bottom}). The errors are $1\sigma$. The orbital period $P_\text{orb} = 2.8173\pm0.00001$~d is taken from \citet{Orosz2001}.}\label{fig:v4641_lc} 
\end{figure}

\begin{figure}
    \centering
    \includegraphics[width=0.95\linewidth]{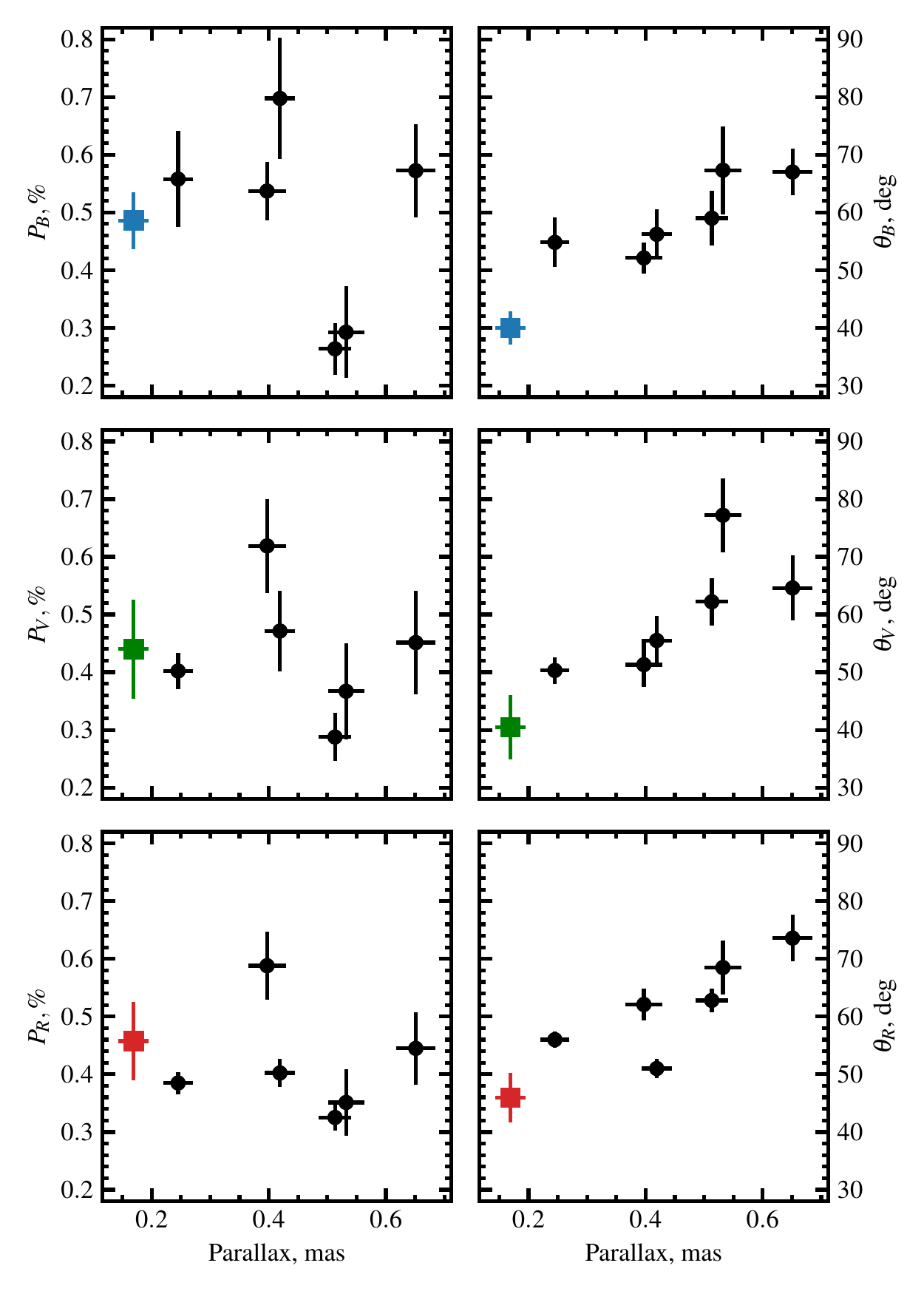}
    \caption{Dependence of the observed polarization degree (left column) and polarization angle (right column) on parallax for \VSGR (coloured squares) and field stars (black circles) in the $BVR$ bands (from top to bottom).}
    \label{fig:v4641}
\end{figure}

\subsection{\VSGR}

Intermediate mass X-ray binary and a microquasar \VSGR shows highly atypical behavior for an X-ray transient.
After the decay of the first outburst \citep{InTZand1999_2}, it underwent a series of over ten failed outbursts (full list can be found in \citealt{Salvesen2020}).
The compact object was dynamically identified as a black hole with the mass $M_\text{bh} = 6.4 \pm 0.6$~\msol, and mass of the companion was found to be $M_{\rm c} \approx 3$~\msol \citep{MacDonald2014}.
The relatively high mass of the companion makes \VSGR one of the largest known Roche lobe-filling X-ray binaries. 
Orbital period of the binary is $P_\text{orb} \approx 2.817$~d \citep{Orosz2001} and the inclination of the orbit $i = 72\degr \pm 4\degr$ \citep{MacDonald2014}.

We observed the source in August 2018 and November 2021 with \DP, when an increased optical activity of \VSGR with $0.3-0.5$~mag enhancement was detected \citep{ATel11949, ATel15014}. 
\DUF\ observations were performed during its quiescent state. 
The object shows rather constant level of polarization $P_\text{obs} = 0.50 \pm 0.05$~per~cent with the position angle of $\theta_\text{obs} \approx 40\degr$ during the whole monitoring period (see Table \ref{tbl:pol}). 
To examine the orbital variability, we folded the polarimetric data with the orbital period -- the resulting polarization shows no dependence on the orbital phase (Fig. \ref{fig:v4641_lc}). 
To analyze the behavior of the IS polarization, we observed six field stars with the parallaxes $\pi = 0.2 - 0.6$~mas (Fig. \ref{fig:v4641}). 
The degree of observed polarization falls in the range of 0.3--0.7~per~cent for all observed stars, while the polarization angle increases with the parallax almost linearly. 
We conclude that the values of average polarization and polarization angle of \VSGR\ in all passbands are consistent with the IS polarization. 
This fact, along with the absence of orbital variability of observed polarization, suggests that \VSGR has no intrinsic optical  polarization in either the quiescence or failed outbursts.

\subsection{\VCYG}

Initially discovered as Nova Cyg 1938, \VCYG  underwent outbursts in 1956 and 1989 \citep{Richter1989}, and two outbursts in 2015 \citep{Barthelmy2015, ATel8453}.
Since its first X-ray detection in 1989 by the \textit{Ginga} satellite \citep{Makino1989}, \VCYG has been extensively monitored in wide energy ranges, including radio, optical, X- and gamma-rays \citep{Corbel2008, Casares1994, Zycki1999, Loh2016}. 
During the 2015 outburst, \VCYG reached 40 Crab in the hard X-rays \citep{Rodriguez2015} and brightened in the optical from $m_V \approx 18$ up to $m_V \approx 11$~mag  \citep{Kimura2016}. 

\VCYG is one of a few low-mass X-ray binaries (LMXB) with orbital parameters and distance known with great accuracy. 
The K-type companion with the mass of $\sim 1$~\msol orbits a $\sim 9$~\msol black hole primary with the orbital period of $\sim 6.5$~days on the orbit, inclined to the observer on $i \approx 67\degr$  \citep{Khargharia2010}. An accurate parallax $\pi = 0.42 \pm 0.02$~mas has been measured by \citet{Miller-Jones2009}.

The 2015 outburst triggered several polarimetric campaigns \citep[e.g.,][]{Shahbaz2016, Tanaka2016, Itoh2017, Kosenkov2017}. 
Optical and near-infrared (ONIR) polarimetric measurements revealed high value of IS polarization ($\sim 7$ per cent, see Table~\ref{tbl:ref_pol} and \citealt{Tanaka2016}) with atypical wavelength dependence -- a potential signature of multiple dust clouds between the source and the observer \citep{Kosenkov2017}.
\VCYG showed statistically significant intrinsic ONIR polarization during its re-brightening (IS polarization was estimated by observing a sample of field stars, see Table \ref{tbl:ref_pol} and \citealt{Kosenkov2017}).
Observed shortly after the outburst, however, \VCYG demonstrated no intrinsic polarization: its observed polarization was identical to the observed polarization of a visually close ($\sim 1\farcs4$, \citealt{Udalski1991}) star, which was reliably resolved as soon as the brightness of the LMXB dropped to the quiescent level.

Several conditions affect the accuracy of polarimetric measurements of \VCYG\ and surrounding field stars.
First, the presence of the visually close companion complicates target separation, especially under poor weather conditions (with seeing $\geq 1 \farcs$). 
Second, relatively high IS extinction ($A_V \approx 3.5$, \citealt{Shahbaz2003}), caused by the proximity to the galactic plane, increases the total integration time needed for reliable measurements, especially in the $B$ filter.
Both new polarimetric measurements of \VCYG made with the \DUF\ suffer from these conditions: the first measurement (made during the technical night allocated for commissioning of \DUF) was too short to reach sufficiently high accuracy in $B$ and $V$ filters, while the second measurement was carried out when the seeing was poor.
Despite these obstacles, the quiescent polarization degree and angle in the $R$ filter (where accuracy is adequate) are in agreement with the polarization obtained for the nearest field stars and are consistent with the previous observations \citep{Kosenkov2017}. 
We therefore see no signs of intrinsic polarization in \VCYG during the quiescence.

\subsection{\maxi}

The LMXB \maxi was discovered in March of 2018 with the Monitor of All-sky X-ray Image (MAXI) nova alert system as a bright X-ray source \citep{Kawamuro2018}, which later was associated with the ASASSN-18ey optical transient \citep{Denisenko2018}.
Over the following $\sim 9$ months \maxi underwent a violent outburst, reaching $m_V \approx 11.5$~mag \citep{Littlefield2018} and $\sim 3$~Crabs in X-rays \citep{ATel11478}.
The initial hard state lasted for $\sim 4$~months and was followed by a soft state, in which the source resided for the same amount of time.
\maxi had never reached the true quiescence after the 2018 outburst has ended; instead, it underwent three \citep{Stiele2020} nearly identical in profile and duration `failed' outbursts, each time increasing its optical brightness from $m_V \approx 18.5$ to $m_V \approx 13.5$~mag.

Since the onset of the 2018 outburst, \maxi was extensively monitored both photometrically and polarimetrically.
Similar to \VCYG, \maxi demonstrated small but statistically significant variable intrinsic optical polarization during rising hard and soft states \citep{Veledina2019,Kosenkov2020b}.
The position angle of intrinsic polarization in the rising hard state \citep[$\sim 24\degr$,][]{Kosenkov2020b} was found to be in good agreement with the position angles of radio \citep{Bright2018} and X-ray \citep{Espinasse2020} jets, providing evidence for a connection between the scattering medium and the jet axis.

\maxi showed no significant intrinsic polarization near the peaks of two failed outbursts (Table \ref{tbl:pol}).
Its observed polarization remained in good agreement with the IS polarization measured from a sample of field stars.
Surprisingly, a dramatically different polarization picture was observed in the (near-)quiescent state: \maxi showed substantially higher (up to 5~per~cent in $B$) intrinsic polarization with polarization angles offset from the jet axis \citep{Poutanen2022}.
The misalignment and large polarization remained surprisingly stable between failed outbursts, suggesting a strong connection to geometrical properties of the source, which can be probed only during inactive phases, otherwise remaining completely obscured by the accretion-ejection processes happening during outbursts.

\subsection{MAXI~J0637--430}

\maxitwo was discovered on 2019 November 2 by MAXI X-ray monitor \citep{Negoro2019}. 
A few hours after the discovery, the optical counterpart with the brightness of $m_u \approx 15$~mag was found in the direction on the X-ray transient with \textit{Swift}/UVOT \citep{Kennea2019}.
Follow-up optical spectroscopic \citep{Strader2019} and X-ray \citep{Tomsick2019} observations suggested that the source is an LMXB hosting a black hole. 
The mass of the compact object has not been reliably measured yet (using, e.g., quiescent state spectroscopy), but it was estimated $M_\text{BH} = 5-12$~\msol  from the X-ray flux and the distance constraint of $d < 10$~kpc \citep{Jana2021}. 

The absence of a reliable estimate of the distance to the object complicates the estimation of IS polarization.
To constrain it, we observed three field stars near \maxitwo with distances in the range of $0.9 - 1.5$~kpc (Table~\ref{tbl:ref_pol}).
The polarization is higher for more distant sources reaching in the $B$-band about 0.6 per cent. 
The observed polarization of \maxitwo (Table \ref{tbl:pol}) is consistent with zero with the 3$\sigma$ upper limit of 1.1 per cent (obtained with Monte Carlo simulations) and is consistent with the IS values. 

\subsection{XTE~J2012+381}

An X-ray transient \XTEtwo was discovered in 1998 by \textit{RXTE} \citep{Remillard1998} reaching 150 mCrab in the 3--20~keV X-ray band. 
The Karl G. Jansky Very Large Array \citep[VLA,][]{Thompson1980} observations obtained in the same year revealed a radio source in the direction of the transient \citep{Hjellming1998}.
Optical observations were able to identify a faint ($m_R \approx  20$~mag) optical counterpart at the coordinates, consistent with the radio and X-ray counterparts \citep{Hynes1999}. 

The faint optical counterpart of \XTEtwo is heavily blended with the visually close ($\sim 1\farcs1$, \citealt{Hynes1999}) and much brighter foreground star. We measured the polarization of the binary ($P_B = 0.06 \pm 0.14$, $P_V = 0.12 \pm 0.09 $, $P_R = 0.17 \pm 0.07$~per~cent), but our observational capacities did not allow us to separate the contribution of \XTEtwo from the contribution of the foreground star to the resulting value of linear polarization. Nevertheless, we obtained the polarization of two nearby field stars and estimated the value of IS polarization in the direction to the binary (Table \ref{tbl:ref_pol}), which can be used in the future polarization studies. 

\begin{table}
    \centering
    \caption{Intrinsic polarization measurements of the observed sample. Both detected values and the upper limits are given. Intrinsic polarization estimate for \XTEtwo\ is not reliable, because of the confusion with the foreground star, and is not given in the table.}
    \label{tab:pol_intr}
    \begin{tabular}{lcccc}
        \hline
        \hline
        Source & State & $P_B$& $P_V$& $P_R$ \\
          &   &  per cent  &    per cent & per cent \\
        \hline
        {\XTEone}  & Q      & $1.2 \pm 0.8 $  & $1.4 \pm 0.8 $  & $1.3 \pm 0.5 $ \\
        {\SwiftJ}  & RH$^a$ & $\leq 0.5 $     & $\leq 0.4 $     & $\leq 0.4 $ \\
        {\fourU}   & S      & $\leq 0.2 $     & $\leq 0.3 $     & $\leq 0.3 $ \\
        {\VSGR}    & RH$^a$ & $\leq 0.1 $     & $\leq 0.1 $     & $\leq 0.1 $ \\
                   & Q      & $\leq 0.1 $     & $\leq 0.1 $     & $\leq 0.1 $ \\
        {\VCYG}    & RH     & $0.8 \pm 0.3 $  & $1.1 \pm 0.1 $  & $0.5 \pm 0.1 $ \\
                   & Q      & $\leq 0.5 $     & $\leq 0.5 $     & $\leq 0.5 $ \\
        {\maxi}    & RH1    & $0.28\pm 0.01 $ & $0.36\pm 0.01 $ & $0.30\pm 0.01 $ \\
                   & RH2    & $0.34\pm 0.02 $ & $0.51\pm 0.02 $ & $0.53\pm 0.02 $ \\
                   & S      & $0.16\pm 0.01 $ & $0.15\pm 0.01 $ & $0.02\pm 0.01 $ \\
                   & DH     & $0.06\pm 0.04 $ & $0.13\pm 0.06 $ & $0.09\pm 0.04 $ \\
                   & RH$^a$ & $\leq 0.3 $     & $\leq 0.4 $     & $\leq 0.3 $  \\
                   & Q      & $3.2\pm 0.2 $   & $1.9\pm 0.2 $   & $0.9\pm 0.1 $ \\
        {\maxitwo} & S      & $\leq 0.2 $     & --               & -- \\
        \hline
    \end{tabular}
    \begin{tablenotes}
      \item Notes:  $^a$Failed outburst. States: Q -- quiescence, RH -- rising hard, 
      \item S -- soft, DH -- decaying hard (see Table~\ref{tbl:pol}).
    \end{tablenotes}
\end{table}

\section*{Summary and discussion}

We performed optical polarimetric observations of a set of Galactic BHXRBs in various spectral states.
Our survey consists of both long- and short-period systems located at low and high Galactic latitudes and residing in quiescent, hard, and soft states.
We used observations of the nearby field stars to estimate the IS polarization in the direction of the selected BHXRBs.
This allowed us to constrain the intrinsic polarization in these sources.
For virtually all systems in our sample, we were able to only put upper limits on the intrinsic polarization -- see summary in Table~\ref{tab:pol_intr}.

Optical and infrared emission of BHXRBs consists of the contributions of several components -- the companion star, accretion disc, inner accretion flow, hot spot/line and jet.
Their relative role in the total spectrum changes with state.
All of them can be polarized, but the polarization degree and its spectral dependence are expected to be different and can be used to discriminate between them.

In the soft state, the optical emission is likely dominated by the disc emission, and the polarization may arise from the scattering processes in its atmosphere.
In the case of pure electron scattering \citep{Cha60,Sobolev1963}, the polarization is expected to increase with the inclination of the disc, reaching a maximum of $P=11.7$~per~cent for the edge-on disc.
The observed soft-state sources, \fourU, \maxitwo and \maxi, on the other hand, show polarization below $\sim1$ per cent, albeit the latter having high inclination.
This may indicate either the complex structure of the accretion disc, such as warp, or the interplay of the scattering and absorption effects in the atmosphere, both of these effects tend to decrease the total polarization.

Likewise, hard-state sources during both regular and failed outbursts show low levels of intrinsic polarization ($<1$~per~cent).
Only MAXI~J1820+070 -- and only during the rising hard state -- has a reliable estimate of intrinsic polarization ($P\sim0.5$~per~cent). 
Its polarization angle coincides with the position angle of discrete ejections detected in the source and the epochs of polarization detection coincide with the detection of winds in the source \citep{Kosenkov2020b}.
This, combined with the red polarization spectrum, may indicate that the polarization is produced by scattering in the wind of the seed photons with the red spectrum.
Such synchrotron emission is produced either in the hot accretion flow or jet.
The absence of significant intrinsic polarization in all hard-state sources in our sample advocates against significant contribution of the jet synchrotron emission itself, as it is expected to be polarized at the level of tens of percent \citep{RadiationProcesses, Veledina2019}.

The detection of a significant, $P_B\sim5$~per~cent, quiescent-state polarization with blue spectrum in \maxi\ put tight constraints on its origin \citep{Poutanen2022}.
Such polarization can be produced by the single Compton scattering in a hot medium, with seed photons coming from the surrounding disc (ring) of a cool matter. 
At the same time, scattering in the disc itself is excluded based on the high value of the polarization, while polarization of jet synchrotron emission is disfavoured based on its spectrum.
In contrast to \maxi, the other two quiescent-state binaries in our sample show low levels of polarization, $P\lesssim 1$~per~cent.
This may indicate the absence of the hot medium in these sources and may indicate that they have entered the true quiescent state (while \maxi\ is still accreting at a very low level).
Future high-precision polarimetric observations of sources in quiescent (and near-quiescent) states are required to confirm the proposed scenario.
Confirmation of the presence of the hot accretion flow in sources undergoing frequent outbursts may indicate its connection to the outburst triggers.

\section*{Acknowledgements}

DIPol-2 and DIPol-UF polarimeters is a joint effort between University of Turku and Leibniz Institut f\"{u}r Sonnenphysik. 
Polarimetric observations with DIPol-UF were performed at the Nordic Optical Telescope, owned in collaboration by the University of Turku and Aarhus University, and operated jointly by Aarhus University, the University of Turku, and the University of Oslo, representing Denmark, Finland, and Norway, the University of Iceland and Stockholm University at the Observatorio del Roque de los Muchachos, La Palma, Spain, of the Instituto de Astrofisica de Canarias. We are grateful to the Institute for Astronomy, University of Hawaii for the observing time allocated for us on the T60 telescope. 
A.V. acknowledges support from the Academy of Finland grant 309308. 
V.K. thanks Vilho, Yrj\"o and Kalle V\"ais\"al\"a Foundation,
A.V. and J.P. received funding from the Russian Science Foundation grant 20-12-00364.  


\section*{Data Availability}
The data underlying this article will be shared on reasonable request to the corresponding author.




\bibliographystyle{mnras}
\bibliography{lmxb_pol} 





\bsp	
\label{lastpage}
\end{document}